\documentclass[5p, longtitle, authoryear]{elsarticle}
\usepackage[dvipsnames]{xcolor}
\usepackage[normalem]{ulem} % to striketrhourhg text
\usepackage{bm}
\usepackage{multirow}
\usepackage{caption}
\usepackage{subcaption}
\usepackage{xspace}
\usepackage{amsmath}
\usepackage[switch]{lineno}
\usepackage{hyperref}
\hypersetup{
    colorlinks=true,
    linkcolor=blue,
    citecolor=blue,
    urlcolor=blue,
}

\newcommand{\python}{\texttt{python}\xspace}

\newcommand{\gammapy}{\texttt{Gammapy}\xspace}
\newcommand{\cpp}{\texttt{C++}\xspace}
\newcommand{\rootcern}{\texttt{ROOT}\xspace}
\newcommand{\mars}{\texttt{MARS}\xspace}
\newcommand{\melibea}{\texttt{melibea}\xspace}
\newcommand{\magicconverter}{\texttt{magic\_dl3}\xspace}
\newcommand{\automagic}{\texttt{automagic}\xspace}
\newcommand{\fits}{\texttt{FITS}\xspace}
\newcommand{\cfitsio}{\texttt{CFITSIO}\xspace}
\newcommand{\github}{\texttt{GitHub}\xspace}
\newcommand{\jointcrab}{\texttt{joint-crab}\xspace}
\newcommand{\he}{HE\xspace}
\newcommand{\vhe}{VHE\xspace}
\newcommand{\fermi}{\textit{Fermi}-LAT\xspace}
\newcommand{\magic}{MAGIC\xspace}
\newcommand{\hess}{H.E.S.S.\xspace}
\newcommand{\veritas}{VERITAS\xspace}
\newcommand{\fact}{FACT\xspace}
\newcommand{\hawc}{HAWC\xspace}
\newcommand{\ctao}{CTAO\xspace}
\newcommand{\gadf}{GADF\xspace}
\newcommand{\mwl}{MWL\xspace}
\newcommand{\irf}{IRF\xspace}
\newcommand{\irfs}{IRFs\xspace}
\newcommand{\sed}{SED\xspace}
\newcommand{\crab}{Crab Nebula\xspace}
\newcommand{\mrk}{Mrk421\xspace}
\newcommand{\ic}{IC310\xspace}
\newcommand{\qso}{QSO B0218+357\xspace}
\newcommand{\ebl}{EBL\xspace}
\newcommand{\diff}{\mathrm{d}}

\journal{Journal of High Energy Astrophysics}

\begin{document}
%\linenumbers
%\titlerunning{Standardised data formats and open-source analysis tools for the \magic telescopes}
%\maketitle

\begin{frontmatter}

\title{Standardised formats and open-source analysis tools for the \magic telescopes data}

\author[1]{The MAGIC Collaboration: S.~Abe}
\author[2]{J.~Abhir}
\author[3]{A.~Abhishek}
\author[4]{V.~A.~Acciari}
\author[5]{A.~Aguasca-Cabot}
\author[6]{I.~Agudo}
\author[7]{T.~Aniello}
\author[8,41]{S.~Ansoldi}
\author[7]{L.~A.~Antonelli}
\author[9]{A.~Arbet Engels}
% ...

\address[1]{Japanese MAGIC Group: Institute for Cosmic Ray Research (ICRR), The University of Tokyo, Kashiwa, 277-8582 Chiba, Japan}
\address[2]{ETH Z\"urich, CH-8093 Z\"urich, Switzerland}
\address[3]{Universit\`a di Siena and INFN Pisa, I-53100 Siena, Italy}
\address[4]{Institut de F\'isica d'Altes Energies (IFAE), The Barcelona Institute of Science and Technology (BIST), E-08193 Bellaterra (Barcelona), Spain}
\address[5]{Universitat de Barcelona, ICCUB, IEEC-UB, E-08028 Barcelona, Spain}
\address[6]{Instituto de Astrof\'isica de Andaluc\'ia-CSIC, Glorieta de la Astronom\'ia s/n, 18008, Granada, Spain}
\address[7]{National Institute for Astrophysics (INAF), I-00136 Rome, Italy}
\address[8]{Universit\`a di Udine and INFN Trieste, I-33100 Udine, Italy}
\address[9]{Max-Planck-Institut f\"ur Physik, D-85748 Garching, Germany}
% ...
\address[41]{also at International Center for Relativistic Astrophysics (ICRA), Rome, Italy}
% ...

\begin{abstract}
% Context
Instruments for gamma-ray astronomy at Very High Energies ($E>100\,{\rm GeV}$) have traditionally derived their scientific results through proprietary data and software. Data standardisation has become a prominent issue in this field both as a requirement for the dissemination of data from the next generation of gamma-ray observatories and as an effective solution to realise public data legacies of current-generation instruments. Specifications for a standardised gamma-ray data format have been proposed as a community effort and have already been successfully adopted by several instruments.
% Aims
We present the first production of standardised data from the Major Atmospheric Gamma-ray Imaging Cherenkov (\magic) telescopes. We converted $166\,{\rm h}$ of observations from different sources and validated their analysis with the open-source software \gammapy. 
% Methods
We consider six data sets representing different scientific and technical analysis cases and compare the results obtained analysing the standardised data with open-source software against those produced with the \magic proprietary data and software. Aiming at a systematic production of \magic data in this standardised format, we also present the implementation of a database-driven pipeline automatically performing the \magic data reduction from the calibrated down to the standardised data level.
% Results
In all the cases selected for the validation, we obtain results compatible with the \magic proprietary software, both for the manual and for the automatic data productions. Part of the validation data set is also made publicly available, thus representing the first large public release of \magic data.
% Conclusions
This effort and this first data release represent a technical milestone toward the realisation of a public \magic data legacy.
\end{abstract}


\begin{highlights}
\item We present the effort to produce the MAGIC telescopes data in a standardised format;

\item we validate the analysis of the standardised data with the open software Gammapy;

\item we implement a database-driven pipeline to systematically produce standardised data;

\item this work represents a technical milestone towards a public MAGIC Data Legacy; 

\item part of the validation data set is made publicly available.
\end{highlights}


\end{frontmatter}

\section{Introduction}
\label{sec:introduction}
The free exchange of astronomical data gathered at different wavelengths joined the different branches of astronomy in the so-called multi-wavelength (\mwl) domain, demonstrating the scientific potential intrinsic to the study of the emission of sources in different energy ranges. While gamma-ray astronomy at high energies (\he, $ 100\,{\rm MeV} < E < 100\,{\rm GeV}$) adopted the same policy of providing public data and software tools \citep{cgro_archive, agile_data_center, fermi_pdmp}, the scientific activity of very-high-energy (\vhe, $E > 100\,{\rm GeV}$) gamma-ray instruments has been traditionally defined instead by proprietary data and software. The issue of data standardisation in \vhe gamma-ray astronomy emerged in the last decade from the decision to build the future Cherenkov Telescope Array Observatory (\ctao), as an open observatory, sharing its observational time and data with the astronomical community \citep{cta_data_management, cta_data_model}. Compelling as the case for data standardisation might be for the dissemination of future \vhe data, it represents a crucial issue for present instruments as well. The current (third) generation of Imaging Atmospheric Cherenkov Telescopes (IACTs, \citealt{hillas_iact_review, de_naurois}) have already operated for two decades and have accumulated a wealth of data that cannot be fully explored by the restricted groups of scientists operating these instruments. The full scientific exploitation of these data sets, even beyond the decommissioning of the telescopes, ideally calls for the realisation of public archival data releases, or ``data legacies''. The adoption of a common format for these data legacies would not only ease their access and usage by the community, since a legacy data release in its native format would require the corresponding proprietary software tool to be maintained and released as well; it would also allow for the exploration and combination of decades of archival gamma-ray observations. 
\par
In the context of future and current \vhe gamma-ray instruments, the effort to define a standard data format started in the second half of the 2010s with the creation of the \textit{Data Formats for Gamma-ray Astronomy} (\gadf) initiative \citep{gadf_proceeding, evolution_data_formats}. The \gadf is a forum where the specifications of a data format for high-level gamma-ray data are discussed through the \github workflow\footnote{\url{https://github.com/open-gamma-ray-astro/gamma-astro-data-formats}}. In parallel to the \gadf specifications, and likewise motivated by the upcoming \ctao, open-source software for the analysis of gamma-ray data, such as \texttt{ctools} \citep{ctools} and \gammapy \citep{gammapy_paper}, were developed. Their data routines were built in compliance with the \gadf specifications, making it possible for the first time to analyse standardised \vhe gamma-ray data with open-source software. Other noteworthy open-source software initiatives are the publication of the \texttt{EventDisplay} reconstruction and analysis software \citep{eventdisplay, eventdisplay_zenodo} used both for the \veritas (Very Energetic Radiation Imaging Telescope Array System) and the \ctao data analysis; and the open development of the reconstruction software for the Large-Sized Telescope prototype (LST-1, \citealt{cta_lstchain}) of the \ctao. Both software, though operating on proprietary data can produce \gadf-compliant data. The first prototypical sample of data in the standardised \gadf format was publicly released by the High Energy Stereoscopic System (\hess) \cite{hess_dl3_dr1} and was crucial in validating the capabilities of the aforementioned open-source analysis tools \citep{gammapy_validation, ctools_validation}. The First g-Apd Cherenkov Telescope (\fact) also made all of its Crab Nebula observations public in the \gadf format\footnote{\url{https://factdata.app.tu-dortmund.de/}}. \cite{joint_crab}, also known as the \jointcrab project, then demonstrated the possibility of effectively performing combined analyses of gamma-ray data from different instruments adopting the standard data format. Data from \crab observations performed by the \he satellite \textit{Fermi} Large Area Telescope (LAT) and by all the then-operating IACTs were produced in the \gadf format and jointly analysed with \gammapy. Beside the already mentioned public \hess and \fact data, the \veritas, and the Major Atmospheric Gamma-ray Imaging Cherenkov (\magic) telescopes produced GADF-compliant data specifically for this project. The \jointcrab dataset thus represented the first joint release of IACT data to the public. Expanding the adoption of the format further, the High Altitude Water Cherenkov (\hawc) Observatory demonstrated that also arrays of particle detectors could produce their gamma-ray data in the \gadf format \citep{hawc_dl3} and extended the exemplary \jointcrab combined spectrum measurement to 5 orders of magnitude in energy.
\par
While for the \jointcrab demonstration only $40\,{\rm min}$ of \magic data were converted to the \gadf format and publicly released, the Collaboration, with the long-term objective of a public data legacy, initiated the effort to systematically produce observations in this standardised format. The objective of this paper is to present this endeavour and to make a first public release of \magic standardised data. We analyse a total of $166\,{\rm h}$ of observations from five different sources representing different scientific and technical analysis cases. We validate the analysis of these standardised data with the open-source software \gammapy by comparing the results obtained against those produced on the same data sets with the proprietary \magic Analysis and Reconstruction Software \citep[\mars,][]{mars_2009, mars_2013}. In this paper, we focus on the validation of the point-like, or one-dimensional, analysis, in which any spatial extension or morphology of the sources is neglected. \par
This paper is structured as follows: in Sect.~\ref{sec:converter} we describe the software we developed to convert the \magic data from their native format to the standardised one. In Sect.~\ref{sec:datasets} we detail the observations used for this study. In Sect.~\ref{sec:validation} we illustrate the different analyses performed with \gammapy on the standardised data and the comparison with the corresponding results obtained with \mars. In Sec.~\ref{sec:data_availability}, we provide public access to part of the dataset adopted in the validation. We conclude in Sect.~\ref{sec:conclusion} by providing some perspectives on the future data legacy of the \magic telescopes. In \ref{sec:likelihood_results_comparison} and ~\ref{sec:flux_detailed_comparison}, we present detailed comparisons of the result of the flux estimation with the two software and investigate the differences produced by the different algorithms adopted.

\section{Conversion of \magic data to the standardised format}
\label{sec:converter}

\begin{figure*}
\centering
    \includegraphics[width=18cm]{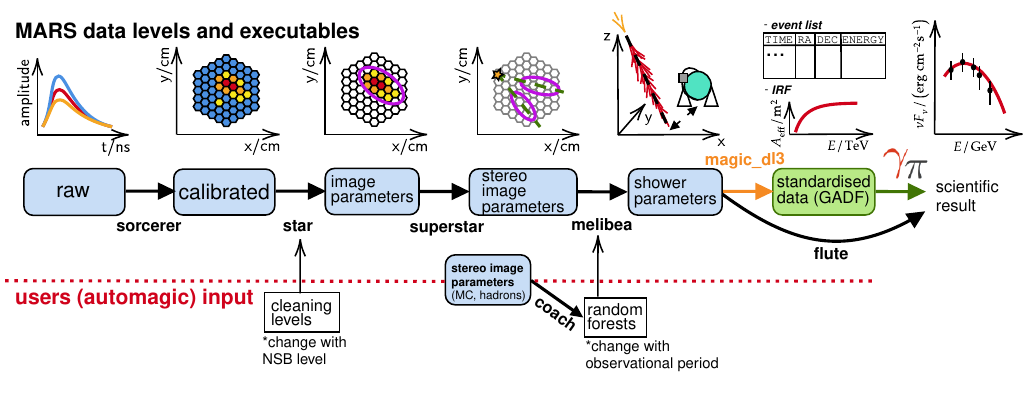}
    \caption{Schematic view of the MAGIC data reduction from raw data to final scientific results. Blue blocks illustrate the different \mars proprietary data levels, black arrows connecting them represent the \mars proprietary executables reducing them. Data levels take the name of the executable producing them (indicated on the arrow). White boxes are used to represent the data-reduction parameters and inputs provided manually by the user or automatically by the \automagic pipeline. Green boxes and arrow represent open data and software. The production of \gadf-compliant data starts from \melibea files (containing the reconstructed shower parameters) using the library developed for this paper (orange arrow). Open-source software such as \gammapy can be then used to extract scientific results from the standardised data.}
    \label{fig:magic_data_reduction}
\end{figure*}

As proposed in \citet{cta_data_model}, one could schematise the progressive data reduction performed by gamma-ray instruments in five different data levels with level 0 representing the raw output of the data acquisition and level 5 scientific results, for example the estimation of the gamma-ray flux of a source. Aiming at facilitating reproducibility and data combination, the \gadf provides specifications for the data level 3 (DL3), containing detector- and calibration-independent information that can be directly used to perform a statistical analysis. DL3 data contain two components: a list of events classified as gamma rays with their estimated coordinates, energies and arrival times and a parametrisation of the response of the system, the so-called instrument response function (\irf), necessary to transform the detector information (e.g. counts) into physical quantities (e.g. fluxes). Building on the experience of the public \fermi data\footnote{\url{https://fermi.gsfc.nasa.gov/ssc/data/analysis/documentation/Cicerone/Cicerone_Data/LAT_DP.html}}, and following the format recommendations for high-energy astrophysical data by NASA's High Energy Astrophysics Science Archive Research Center (HEASARC) \footnote{\url{https://heasarc.gsfc.nasa.gov/docs/heasarc/ofwg/ofwg_recomm.html}}, the \gadf specifications are defined for the Flexible Image Transport System \citep[\fits,][]{fits_1981} file format. 

\subsection{\magicconverter}
\label{sec:magic_dl3}
The \magic proprietary analysis and reconstruction software, \mars, is a \cpp library built on \rootcern \citep{root} that provides several executables performing the different data-reduction steps, schematised in Fig.~\ref{fig:magic_data_reduction}, from raw data to scientific results. Though each data level can be serialised using the \rootcern file format, none of them exactly corresponds to the \gadf DL3 specifications\footnote{Notice that the \mars executable used to produce scientific results, \texttt{flute} (Fig.~\ref{fig:crab_counts_comparison}), directly performs a data reduction (see Sect.~\ref{sec:spectral_analysis}), without storing the event list information. Additionally, it can compute \irf components only for MCs corresponding to a specific observation configuration (see the discussion in Sect.~\ref{sec:crab_multi_offset_dataset}).}. Thus, we developed a proprietary library, \magicconverter, extracting the requested information from the \mars files and storing it in the \gadf-compliant format. \magicconverter is a \cpp library built on \rootcern, \mars, and \cfitsio \citep{cfitsio} that operates on the \mars \melibea data level (see Fig.~\ref{fig:magic_data_reduction}). \melibea files contain reconstructed shower parameters and an estimate of the energy, direction, and arrival time of the primary particle of a set of observed (or simulated) events. Along with these quantities, they specify, per each event, the score of a particle identification algorithm \citep[random forest,][]{magic_rf}. \melibea data produced from observations are used to extract the first component of the DL3 files, that is the event list, by applying a cut on the random forest score to select a list of ``gamma-like'' events. The second component of the DL3 files, the \irf, is composed of different functions representing the collection area and the probability distributions of the energy and direction estimators \citep[see Sec.\ref{sec:spectral_analysis} and][for more details]{magic_performance}. Histograms representing these \irf components are built from Monte Carlo (MC) events, after applying the same selection cuts adopted for the event list selection. As the \mars data reduction down to the \melibea data level is still necessary to produce DL3 data, in the next section we introduce a tool developed to simplify and automatise the systematic generation of DL3 files.

\subsection{\automagic}
\label{sec:automagic}
Different observing conditions can easily multiply the number of parameters (image cleaning settings, data selection and analysis cuts) and input data sets necessary to perform a \mars data reduction. Different MC gamma-ray simulations are produced according to significant changes in the hardware configuration of the telescopes (roughly one per year). Simulated gamma rays, along with observations containing no gamma-ray signal (commonly referred to as hadrons)\footnote{While MC gamma rays are readily available to train the classification algorithm, in order to save the extensive computational time of hadronic shower simulations, hadrons data are manually selected by users examining other observations with no gamma-ray signal that match the observing conditions of the data of interest.}, are used to train the particle identification algorithm. This implies, for analysers considering a data set spanning several years, the necessity to train a number of different algorithms, one per each of the hardware configuration (i.e. MC productions) being considered. While the data reduction would normally start from already computed stereo image parameters, \magic developed a procedure to analyse data taken under different moon conditions \citep[described in][]{magic_performance_moonlight}. The latter requires users to start the analysis from the calibrated data level (see Fig.~\ref{fig:magic_data_reduction}) and to perform the cleaning of the shower images with parameters tuned to suppress the pixels illuminated by the diffuse night-sky background (NSB). In case of data taken under different NSB conditions, not only the observations, but also the MC and the hadrons shower images require this tuned cleaning. Therefore the analysis of a data set characterised by different MC productions and NSB conditions requires analysers to handle hundreds of ${\rm GB}$ or ${\rm TB}$ of data, and perform a cumbersome number of different low-level data reduction steps. To orchestrate these different processes, we developed \automagic, a database-supported \python pipeline that automatically performs all the data reduction steps from calibrated down to \melibea and DL3 files. The project was started with the aim to simplify individual analyses, but was soon adopted by the \magic Collaboration to systematically process, in a reproducible and automatic way, large volumes of data and directly produce high-level data sets. The pipeline runs in the computing cluster of the \magic Data Center, at the Port d'Informaci{\'o} Cient{\'i}fica (PIC)\footnote{\url{https://www.pic.es/}}. Its input consists simply of a source name, a given time period, and the data selection cuts (e.g. zenith range, atmospheric transmission, etc.). Having access to the whole archive of \magic data, \automagic localises, in the PIC file system, the calibrated data corresponding to the user selection and creates database tables representing the data files at different stages of data reduction (down to DL3). The pipeline identifies the MC periods within the selected data set and automatically selects the appropriate hadrons and MC data to train the corresponding classification algorithms. It additionally classifies the NSB levels of the observations and configures the appropriate settings for the image cleaning (see Fig.~\ref{fig:magic_data_reduction}). Having identified all the input data and the configurations for the different data reduction steps, \automagic runs them sequentially in the PIC computing nodes. By parallelising the execution of the individual reduction steps and by creating database tables to also track the status of the different jobs, \automagic is able to manage large data volumes and to perform a reproducible and resource-efficient data reduction. The \automagic pipeline is proprietary to the Collaboration, as it is designed to work within the \magic Data Center at PIC.

\section{Data samples selected for the validation}
\label{sec:datasets}
Having implemented the tools to produce standardised \gadf-compliant DL3 data, we aim to validate that their analysis produces results consistent with those obtained with the \magic proprietary software. For the analysis of the standardised data we adopt the open-source software \gammapy \citep[\texttt{v1.1},][]{gammapy_v1_1}, the software tool that will also form the basis of the future \ctao analysis software. 
%This work provides the first systematic validation of a large \magic data set, corresponding to $\bm{166\,{\rm h}}$ of observations, in the \gadf-compliant format. 
In this paper, we focus on the validation of the point-like or one-dimensional analysis: signal and background events are extracted from fixed regions in the sky and their distribution as a function of the energy is fitted. This analysis case, suited for individual sources with extension below the instrument's point spread function, is appropriate for most of the \magic observations. To test the use cases most commonly considered in a one-dimensional analysis, we choose $6$ data sets representing different scientific and technical scenarios. They are presented in Table~\ref{table:datasets} and described in detail in what follows.

\begin{table*}
\caption{\magic data sets used for the DL3 data validation.}
\label{table:datasets}
\centering
\begin{tabular}{l l l l l l l}
\hline\hline
source & period & dark & obs. time & science case & technical case & size (n. of runs)\\
\hline
\crab & 2011-2012 & yes & $42\,{\rm h}$ & bright steady source & different offsets & $45\,{\rm MB}\;(169)$\\
\crab & 2018-2019 & no & $20\,{\rm h}$ & bright steady source & different moon conditions & $14\,{\rm MB}\;(120)$\\
\ic & 2012 & yes & $3.5\,{\rm h}$ & bright hard source & different offsets & $3\,{\rm MB}\;(11)$\\
B0218+357 & 2014 & yes & $2\,{\rm h}$ & dim soft source & - & $2\,{\rm MB}\;(7)$\\
\mrk & 2014 & yes & $42\,{\rm h}$ & bright variable source & - & $29\,{\rm MB}\;(176)$\\
M15 & 2015-2016 & yes & $57\,{\rm h}$ & non detection & - & $47\,{\rm MB}\;(200)$\\
\hline
\end{tabular}
\end{table*}

\subsection{\crab}
The \crab represents a reference source for \vhe gamma-ray astronomy, being the brightest steady source emitting above hundreds of ${\rm GeV}$ \citep[exceptional gamma-ray flares have been detected only at \he, see][]{crab_flares_fermi, crab_flares_agile}. Indeed, it was the first source to be detected at \vhe by \cite{crab_whipple} \citep[see also][for an overview of the source and its emission mechanisms]{crab_review}. We select two different samples of \magic observations of the \crab related to two different technical aspects of the standardised data production we aim at validating.

\subsubsection{Observations at different pointing offsets}
\label{sec:crab_multi_offset_dataset}
The first sample consists of $42\,{\rm h}$ of \crab observations performed between 2011 and 2012, that we consider to validate two scientific results most commonly obtained in a one-dimensional analysis: spectrum and light curve. These data contain the sample of observations used to estimate the \magic performance after the upgrade of the stereoscopic system completed in 2012 \citep{magic_performance}. Most of the \magic observations are conducted in the so-called wobble mode \citep{fomin_1994}, with the telescopes tracking sky coordinates which are typically $0.4^{\circ}$ from the source nominal position. This results in the source having a projected position in the camera plane $0.4^{\circ}$ from its centre, which facilitates the background estimation (see Fig.~\ref{fig:energy_depenent_spectrum_exctraciton}). As this configuration represents most of the observations performed by \magic, MC gamma rays are commonly simulated with arrival directions at $0.4^{\circ}$ from the telescope axis. MCs thus produced are referred to as ``ring-wobble'' simulations, and are adopted for most of the analyses. As the \irf dependency in the \gadf specification is expressed in projected camera coordinates (or offset from the centre, in case of radial symmetry) rather than sky coordinates, we refer to \irfs generated from these MCs as ``single-offset'', since the offset dependency of the \irf components is restricted to a single value ($0.4^{\circ}$). To take into account wobble-mode observations with different pointing offsets, or to study extended sources, ``diffuse'' MCs are produced, not applying the previous restriction on the direction of the simulated events. We refer to \irfs generated from this MC as ``multi-offset'', as the full dependence of the \irf components with the offset from the camera centre is considered (all the \irfs adopted in these tests are radially symmetric in camera coordinates). To test the spectral analysis at several pointing offsets, we first select $30\,{\rm h}$ of \crab observations at the standard $0.4^{\circ}$ offset and produce single-offset IRFs. We then add to the sample $20\,{\rm h}$ of \crab observations with different pointing offsets: $0.20^{\circ}, 0.35^{\circ}, 0.40^{\circ}, 0.70^{\circ}, 1.00^{\circ}, 1.40^{\circ}$, producing multi-offset IRFs. Note that the multi-offset data at $0.40^{\circ}$ are a subset of the $30\,{\rm h}$ single-offset sample, but reprocessed with diffuse MC. The zenith range of the single-offset sample is $[5^{\circ}, 50^{\circ}]$, while that of multi-offset sample is $[5^{\circ}, 35^{\circ}]$. MC gamma rays are generated with an uniform distribution of cosine of zenith and azimuth. In order to account for the zenith and azimuth dependency of the \irf, MCs are re-weighted according to the effective time distribution in zenith and azimuth  of the real events. Proprietary and standardised data sets are composed of observational ``runs'', corresponding to chunks of $15$ to $20\,{\rm min}$ of data acquisition. The number of runs per data set is indicated in Table~\ref{table:datasets}, along with the size of the standardised data. For a comparison with the proprietary format, in the case of the \crab single-offset data set, we converted $21\,{\rm GB}$ of \melibea files, the reduction factor from \melibea to DL3 thus being $\sim 500$.

\subsubsection{Observations under different moonlight conditions}
\label{sec:crab_moon_dataset}
Another sample of \crab observations is used to test the reliability of the high-level data produced by the \automagic pipeline. As already described in Sect.~\ref{sec:automagic}, \automagic is designed to produce \magic DL3 data automatising the process of data selection and reduction, noticeably simplifying cumbersome analyses as those of observations taken under different moon conditions. To test \automagic's capabilities, we consider a sample of $20\,{\rm h}$ of \crab observations with four different moon/NSB illumination conditions. The NSB level was measured in units of the mean direct currents generated in the photo-multiplier tubes of the MAGIC I camera, with ``dark'' observations corresponding to mean currents typically around $1\,\mu A$ \citep[see][]{magic_performance_moonlight}. These moon observations were performed between November 2018 and September 2019, with pointing offset $0.4^{\circ}$ and zenith range $[5^{\circ}, 50^{\circ}]$. We manually perform the data selection and reduction on this data set while, in parallel, let \automagic perform the same processes by specifying as the only input of the pipeline the source name and the start and end dates of the observations to be considered. We finally compare the spectra obtained with the manual and the automatic approaches. 

\subsection{\qso and \ic}
We include the jetted active galactic nuclei \qso and \ic in the validation sample to test the spectrum estimation in case of a soft and hard gamma-ray source, respectively. Given the non-negligible absorption of their gamma-ray emission by the extragalactic background light (\ebl, \citealt{cooray_ebl_review}), we also compare the \ebl absorption treatment in \mars and \gammapy. \qso is a gravitationally-lensed blazar located at redshift $z=0.944$ \citep{b0218_redshift}. We consider data from July 2014 corresponding to the detection in \vhe of the delayed component of a gamma-ray flare first observed by \fermi \citep{magic_b0218_detection}. The source, displaying a very soft gamma-ray spectrum (with measured spectral index $\sim -4$), was observed for $2\,{\rm h}$ in the zenith range $[5^{\circ}, 50^{\circ}]$ with the standard pointing offsets of $0.4^{\circ}$. \ic is a radio galaxy located at a redshift $z=0.0189$ \citep{ic310_redshift}, embedded in the Perseus cluster. We consider the data set corresponding to the November 2012 gamma-ray flare described in \cite{magic_ic310_flare}. This contains the source brightest and hardest gamma-ray emission. \ic was observed for $5\,{\rm h}$ in the zenith range $[5^{\circ}, 35^{\circ}]$ with two different pointing offsets of $0.4^{\circ}$ and $0.94^{\circ}$ (due to a pointing strategy optimised to include at once different sources of the Perseus cluster in the \magic field of view).

\subsection{\mrk}
\label{sec:mrk_dataset}
To test the estimation of the gamma-ray flux of a very variable source, we consider \mrk, one of the brightest blazars observed at all wavelengths \citep{magic_mrk421_2011}. We selected $42\,{\rm h}$ of observations from 2014, already presented in \cite{magic_mrk421_2014}, that we use to test the light curve estimation in the case of a very fast (sub-hour time scale) flux variability. The source was observed with $0.4^{\circ}$ pointing offset, in the zenith range $[5^{\circ}, 70^{\circ}]$.

\subsection{M15}
\label{sec:m15_dataset}
Another computation that can be performed in a one-dimensional analysis is the estimation of upper limits on the flux of a undetected target. For this analysis case, we select $57\,{\rm h}$ of observation of M15, a globular cluster whose potential \vhe emission was investigated by \magic in \cite{magic_m15}. The source was observed with $0.4^{\circ}$ pointing offset, in the zenith range $[5^{\circ}, 50^{\circ}]$.

\section{Validation}
\label{sec:validation}

\begin{figure*}
    \centering
    \includegraphics[width=17cm]{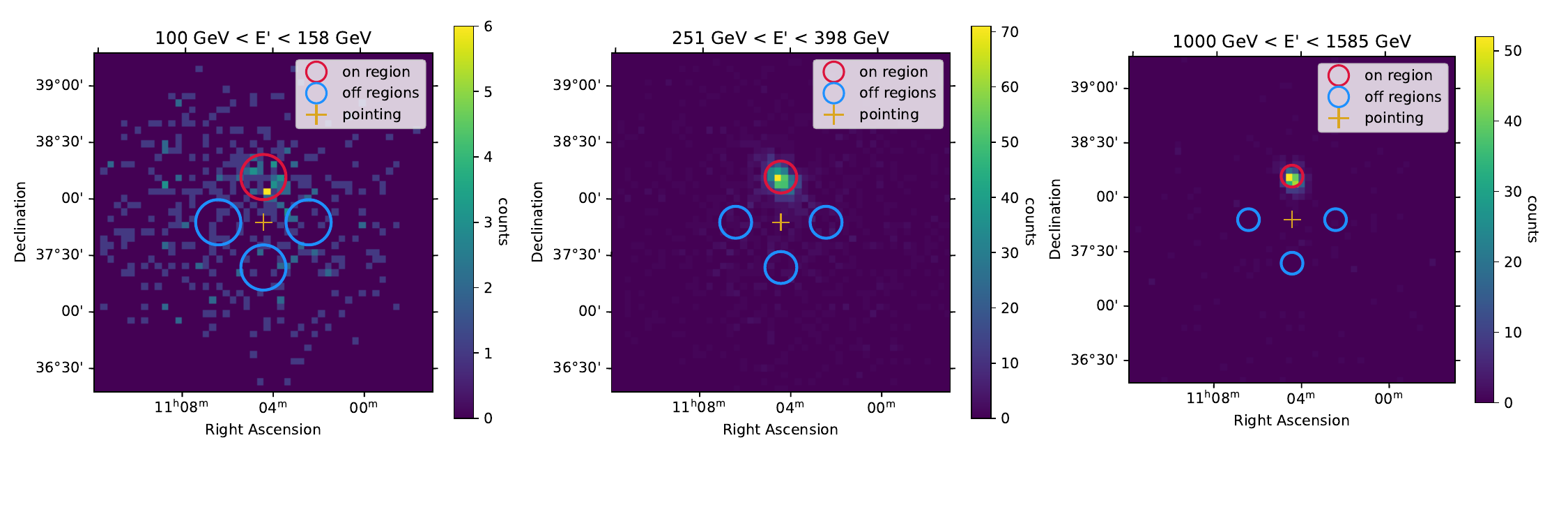}
    \caption{Energy-dependent signal and background estimation. Events from a single run of \mrk observations. The three different panels show the distribution of reconstructed coordinates of gamma-ray candidates in three different bins of estimated energy, in order to illustrate how the size of the on and off regions varies with energy.}
    \label{fig:energy_depenent_spectrum_exctraciton}
\end{figure*}

The process that goes from high-level data, standardised or proprietary, to scientific results (i.e. the last arrow in Fig.~\ref{fig:magic_data_reduction}) typically consists of two parts: a data reduction process that produces binned data from event lists and \irfs\footnote{Unbinned analysis considering the measured quantities of individual events can also be considered, see Sect.~1.2.5 of \cite{gamma_data_analysis_handbook}.}; and a statistical analysis, that uses the latter to estimate fluxes. Validating the results of the analysis of standardised data with \gammapy against \mars therefore implies comparing both data reduction and statistical algorithms. In this section, we provide a brief theoretical background of the algorithms used for reduction and analysis, before moving on to the results of their validation.

\subsection{Point-like analysis}
\label{sec:spectral_analysis}

In the most general case, when analysing gamma-ray observations, one would like to estimate the flux of one or more sources in the field of view (FoV). An analysis accounting for the position and morphology of different sources in the FoV is referred to as spectro-morphological or three-dimensional (two sky coordinates and the energy of the events are the dependencies considered for binning the data and for the emission model to be fitted). In this type of analysis, an estimate of the gamma-ray background (i.e. hadronic showers misidentified as gamma rays) over the whole FoV is required. When considering instead a single source with negligible extension, the background can be directly estimated from the observation itself with aperture photometry techniques \mbox{\citep{berge_2007}}. While the signal is measured from a region referred to as ``on'', centred on the source, the background is commonly extracted from one or more ``off'' regions, often symmetric to the on region with respect to the camera centre, as illustrated in Fig.~\ref{fig:energy_depenent_spectrum_exctraciton}. As this analysis integrates out all spatial dependencies and considers only the distributions of events as a function of energy, it is commonly referred to as ``point-like'' or ``one-dimensional'' analysis. Moreover, the sizes of the on and off regions can be tuned to reflect the improvement of the angular resolution of the telescopes with energy\footnote{The size of the on and off regions can be computed from MC simulations by selecting, across different energy bins, the radius of the region containing a common fraction of events simulated from a point-like source}. As part of this work, this energy-dependent signal and background extraction, illustrated in the different panels of Fig.~\ref{fig:energy_depenent_spectrum_exctraciton}, was implemented in \gammapy and made available since \texttt{v1.0} \citep{gammapy_v020}.
\par
In the one-dimensional analysis, in order to estimate the spectrum of a source, an analytical model is considered to represent the differential flux of the source as a function of true energy 
\begin{equation}
\frac{\diff \phi}{\diff E}(E; \bm{\theta})\,[{\rm cm}^{-2}\,{\rm s}^{-1}\,{\rm TeV}^{-1}],
\end{equation}
where $\bm{\theta}$ is the set of parameters specifying the analytical model. They can be estimated through a statistical procedure, e.g. a likelihood maximisation. In order to do so, as the result of the signal/background estimation is a histogram of counts vs (estimated) energy (see e.g. Fig.~\ref{fig:crab_counts_comparison}), one needs to translate the analytical flux model into ``predicted'' source counts. This is done by folding the assumed differential flux model with the \irf. In case of an observation that can be characterised by a single \irf (i.e. stable observing conditions) the number of gamma-ray events in the $k$-th bin of estimated energy is given by
\begin{equation}
\begin{split}
g_k(\bm{\theta}) &= t_{\rm eff} \int_{E'_{k}}^{E'_{k + 1}} \diff E' \, \int_{0}^{\infty} \diff E \, {\rm IRF}(E' | E) \, \frac{\diff \phi}{\diff E}(E; \bm{\theta}) \\
                             &= t_{\rm eff} \int_{E'_{k}}^{E'_{k + 1}} \diff E' \, \int_{0}^{\infty} \diff E \, A_{\rm eff}(E) \, R(E' | E) \, \frac{\diff \phi}{\diff E}(E; \bm{\theta}),\\
\end{split}
\label{eq:predicted_counts}
\end{equation}
where, $t_{\rm eff}$ is the effective time of the observations and, in the second line, we have factorised the \irf into two components: effective area, $A_{\rm eff}(E)$, and energy dispersion, or migration matrix, $R(E' | E)$, expressing the probability density function of the energy estimator. We have adopted the convention of expressing estimated (measured) quantities as primed. Observed and predicted counts can be combined in a likelihood assuming a Poissonian distribution of the on and off events in each of the $n_{E'}$ bins in estimated energy considered for the analysis
\begin{equation}
\begin{split}
\mathcal{L}(\bm{\theta} &| \{N_{{\rm on}, k}, N_{{\rm off}, k}\}_{k=1,...,n_{E'}}) =\\ 
&\prod_{k=1}^{n_{E'}} {\rm Pois}(g_{k}(\bm{\theta}) + b_{k} | N_{{\rm on}, k}) {\rm Pois}(\alpha\,b_{k} | N_{{\rm off}, k}),\\
\end{split}
\label{eq:likelihood}
\end{equation}
where $N_{{\rm on}, k}$ and $N_{{\rm off}, k}$ are the number of events observed in the on and off regions, respectively, in the $k$-th estimated energy bin. The likelihood is maximised by varying the values of the model parameters, $\bm{\theta}$, and consequently the number of predicted source counts in each energy bin, $g_k$; the number of predicted background counts, $b_k$, is instead commonly treated as a nuisance parameter \citep[more details can be found in Appendix A. of][]{piron_2001}. $\alpha$ is a factor taking into account the different exposures of the on and off regions (e.g. in the case of Fig.~\ref{fig:energy_depenent_spectrum_exctraciton}, $\alpha=3$ as there are three off regions of assumed equal acceptance). To obtain a result from data sets corresponding to different observing conditions (i.e. \irfs) or even different instruments, their likelihood terms as in Eq.~\ref{eq:likelihood} can be factored. 

\subsection{Validation of the data reduction}
\label{sec:data_reduction_comparison}

\begin{figure}
    \centering
    \resizebox{\hsize}{!}{\includegraphics{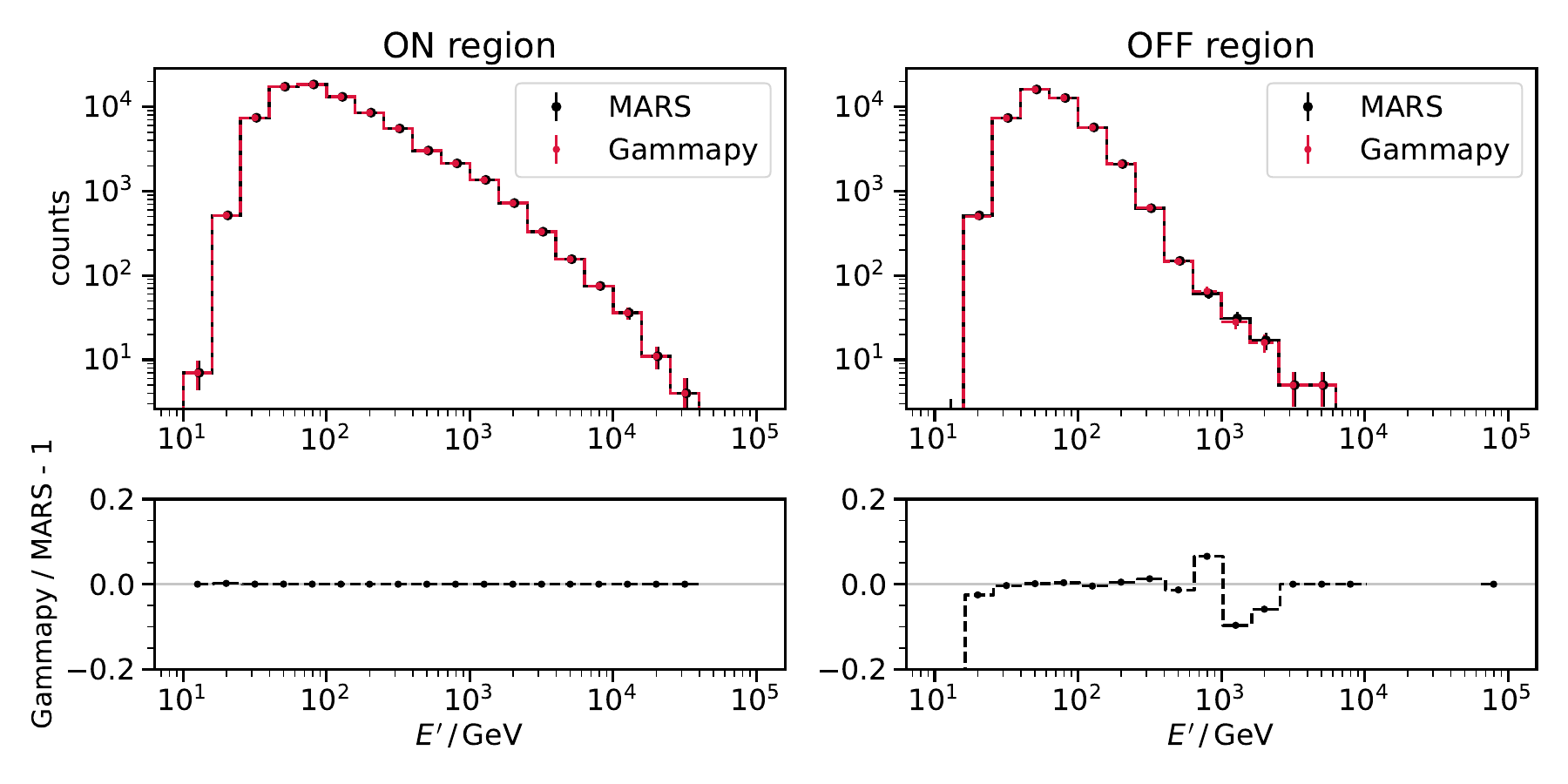}}
    \caption{Comparison of the observed number of counts in the on and off regions with the two software using the proprietary (\mars) and the standardised files (\gammapy), respectively.}
    \label{fig:crab_counts_comparison}
\end{figure}

\begin{figure}
    \centering
    \resizebox{\hsize}{!}{\includegraphics{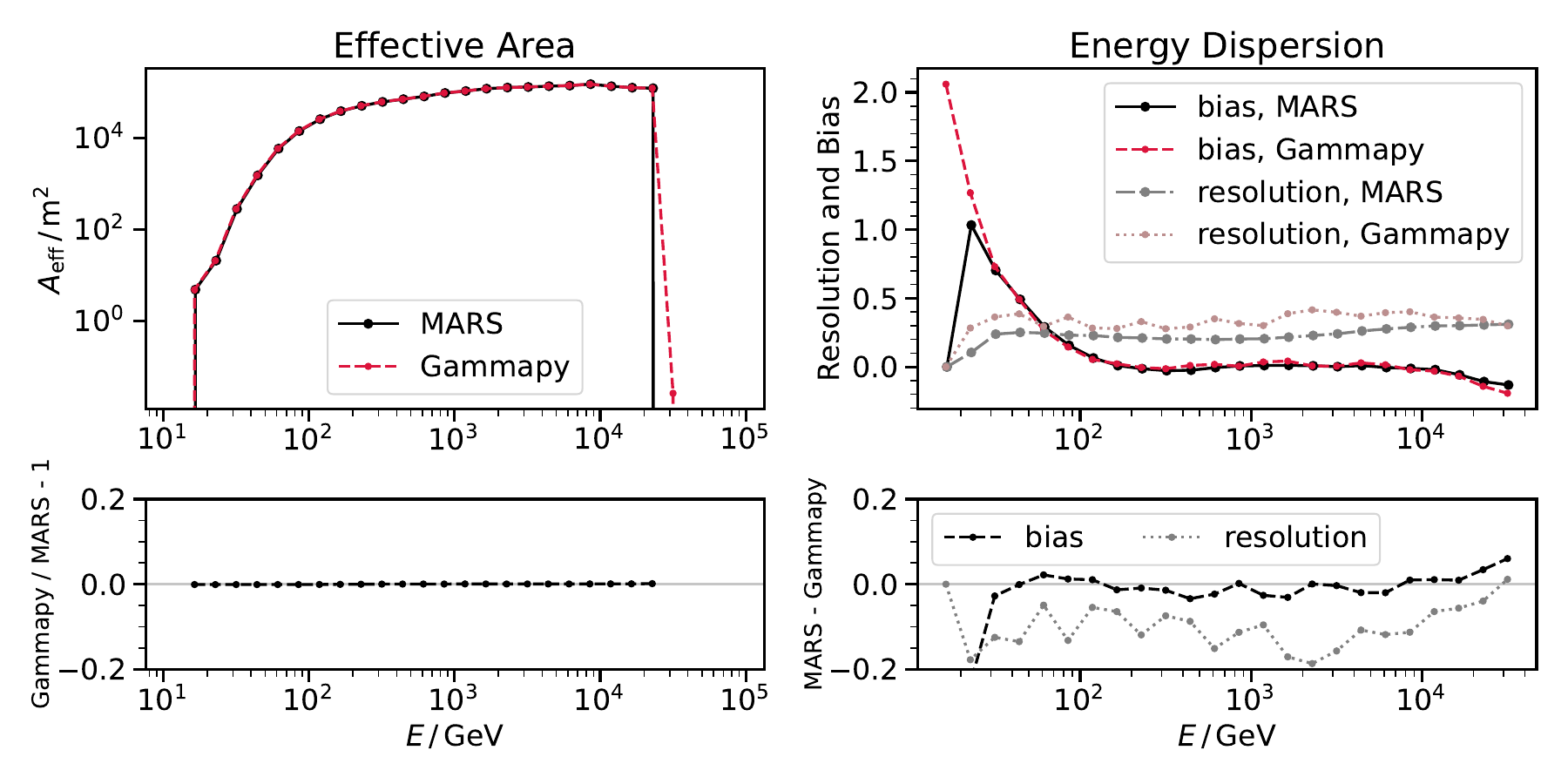}}
    \caption{\textit{Left}: comparison of the effective area stored in the proprietary and standardised files. \textit{Right}: comparison of bias and resolution of the energy dispersion (see Section 4.4 in \citealt{magic_performance} for their definition) stored in the proprietary and standardised files.}
    \label{fig:crab_irf_components_comparison}
\end{figure}

The first step to validate the analysis of standardised data with \gammapy consists in examining the results of the data reduction, that is histograms of the ``on'' and ``off'' counts and of the \irf components\footnote{\irf components are stored in the DL3 files as tables of values of the response at different offsets from camera centre and at different true energies. These tables are interpolated by \gammapy's data routines \citep{table_interpolation} thus obtaining functions that can be evaluated according to the offset and energy binning chosen for a particular analysis.} to be used for statistical analyses (e.g. for the likelihood maximisation previously described). We illustrate, in Fig.~\ref{fig:crab_counts_comparison} and \ref{fig:crab_irf_components_comparison} the on and off counts and the \irf components, respectively, obtained with the standardised and proprietary pipelines. Though we illustrate results only for the single-offset \crab data sample, we observe an almost exact agreement in most of the cases. The difference in the last value of the effective area is an effect of \gammapy's interpolation; differences in the energy dispersion values are due to the slightly different formats adopted by the two pipelines (while a two-dimensional histogram of $E'$ vs $E$ is stored in \mars, a histogram of $E'/E$ vs $E$ is stored in the DL3 files according to the \gadf specs).

\subsection{Validation of the statistical results}
Having validated the data reduction, we move on to examining the statistical results that can be obtained with the different scientific and technical cases in our data samples.

\subsubsection{Spectrum and light curve of the \crab observed at different offsets}
\label{sec:crab_multi_offset_results}

\begin{figure*}
    \centering
    \includegraphics[width=17cm]{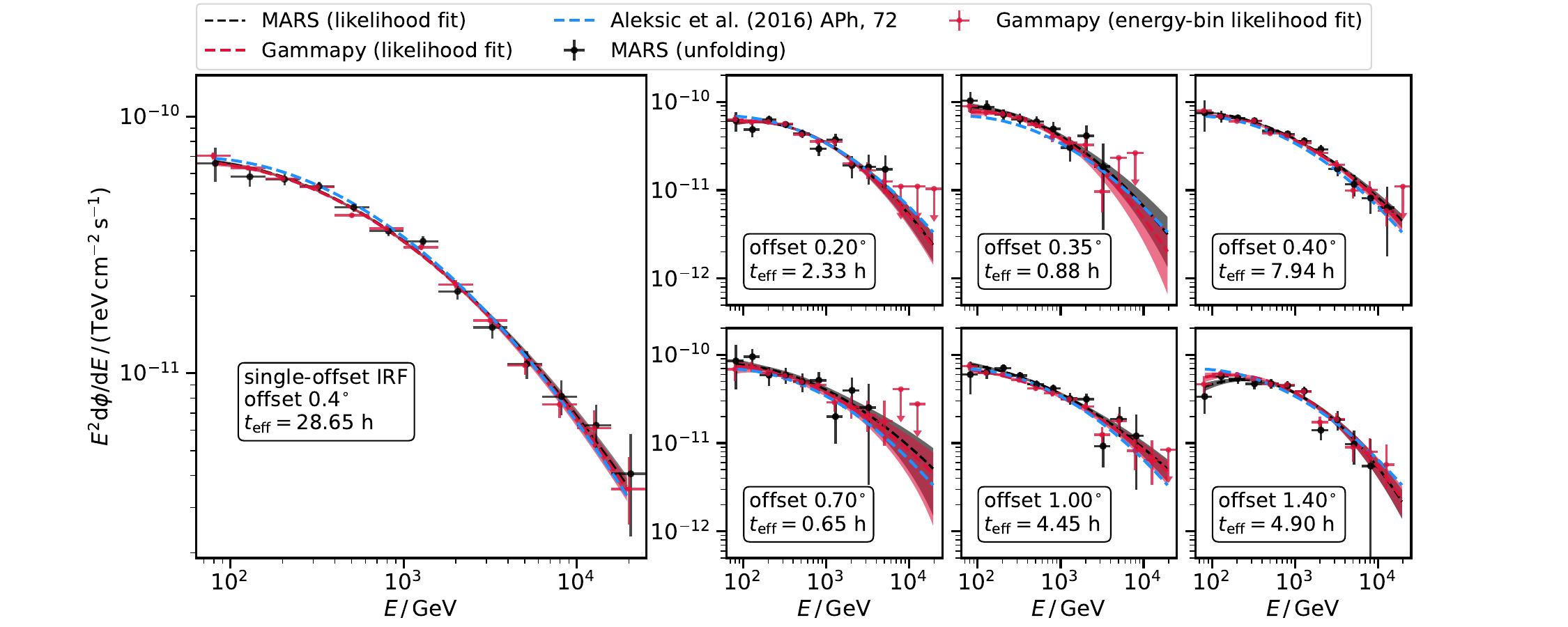}
    \caption{\sed of \crab estimated at different camera offsets and with different MC productions. Dashed lines represent the result of the likelihood fit considering all the events, the bands represent the uncertainty on this results. Points represent a flux estimation considering only the events in a given energy bin, they are computed with different methods by the two software (see main text). The spectrum obtained in \cite{magic_performance} from the same data set is added for reference.}
    \label{fig:crab_sed}
\end{figure*}

\begin{figure*}
    \centering
    \includegraphics[width=17cm]{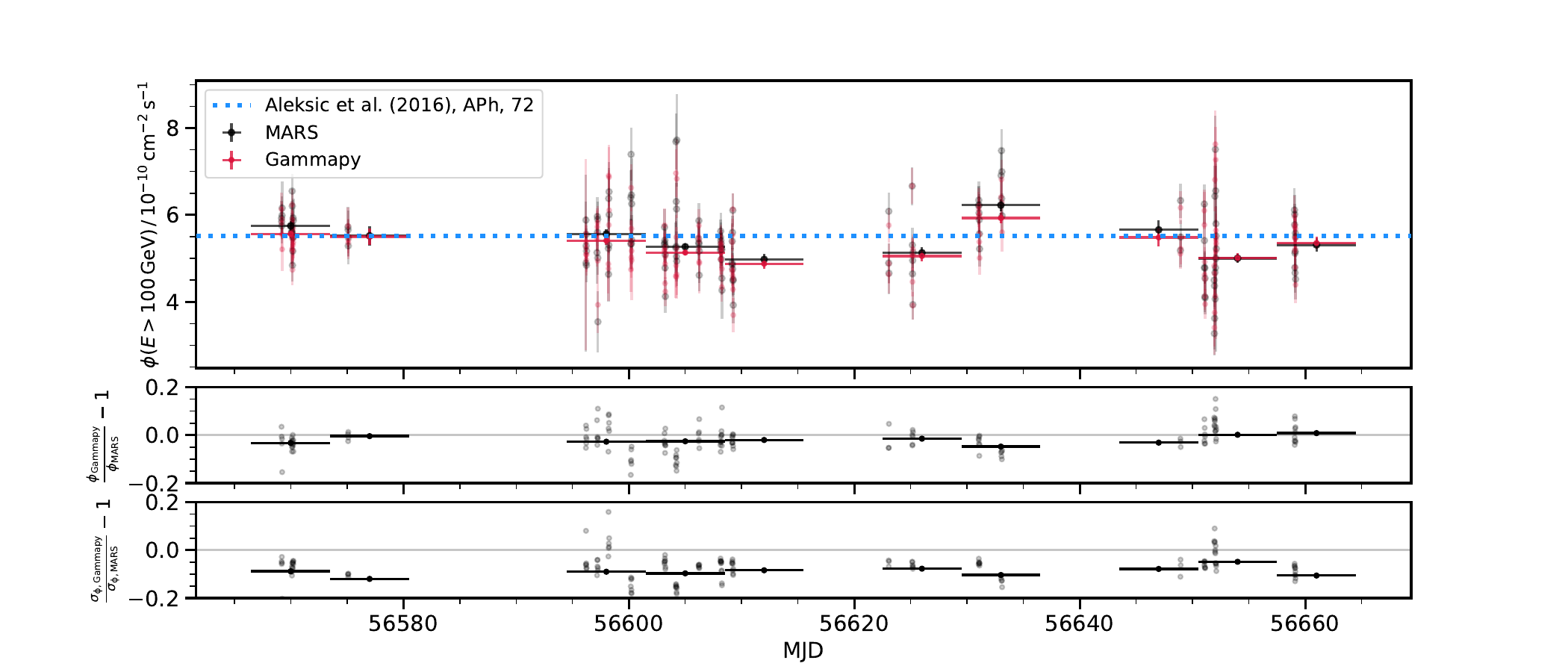}
    \caption{LC of the \crab obtained with the single- and multi-offset data sets. Transparent points correspond to run-wise binning, solid points to weekly binning. The blue dotted lines represents the measurement obtained by \magic in \cite{magic_mrk421_2014}.}
    \label{fig:crab_lc}
\end{figure*}

We use the single- and multi-offset data sets  described in Sect.~\ref{sec:crab_multi_offset_dataset} to estimate the spectrum of the \crab. We assume a log-parabolic spectral model
\begin{equation}
\frac{\diff \phi}{\diff E}(E; \phi_0, \alpha, \beta, E_0) = \phi_0 \, \left( \frac{E}{E_0} \right)^{-\alpha -\beta \log_{10}\left(\frac{E}{E_0}\right)},
\label{eq:log_parabola}
\end{equation}
fixing the arbitrary reference energy $E_0$ to $1\,{\rm TeV}$. We obtain an estimate of the amplitude $\phi_0$, and of the two spectral indices, $\alpha$ and $\beta$, by maximising the likelihood in Eq.~\ref{eq:likelihood}. In Fig.~\ref{fig:crab_sed} we show the spectral energy distribution (SED) obtained from the single- and multi-offset samples and provide a detailed comparison of the estimated spectral parameters and their errors in Fig.~\ref{fig:crab_parameters}. We observe an excellent agreement between the results of the standardised and proprietary pipelines. We consider this result as the main focus of our validation of the spectrum estimation, as \mars and \gammapy implement the same likelihood method. As a secondary check, we also compute spectral data points, or ``flux points''. For this estimation, the methods adopted by the two pipelines differ. \gammapy computes flux points by re-perfoming the likelihood fit considering only the events with energies within the given $k$-th estimated energy bin. Starting from the best-fit model of the broadband fit, only the amplitude $\phi_0$ is re-fitted, thus returning a flux measurement in each energy bin\footnote{We remark that despite the flux points computation is performed considering only the events in a given estimated energy bin, they are represented, in all the plots, as bins in true energy.}. \mars instead performs an unfolding \citep{unfolding_magic} procedure, migrating the excess events in their true energy bins accounting for the energy dispersion. The flux is then directly estimated from the migrated excess events. Despite the different methods adopted, we observe a good agreement also in the estimated flux points. 
\par
We also employed the single- and multi-offset \crab data sets to compute the light curve (LC). In this case, the software implement a different estimation of the integral flux $\phi(E > E_{\rm min})\,/\,[{\rm cm}^{-2}\,{\rm s}^{-1}]$ in a given time bin. \gammapy performs the likelihood maximisation described in Sect.~\ref{sec:spectral_analysis}, considering all the events within a given time bin, and (as for the spectral flux points) re-fitting only the amplitude of the spectrum obtained from the broad-band fit. \mars instead computes the LC flux points directly from the number of excess events, without performing a likelihood maximisation (the actual computation is described in detail in \ref{sec:flux_detailed_comparison}). The LCs obtained with the two software are presented in Fig.~\ref{fig:crab_lc}, where we show both a run-wise (transparent) and a weekly (solid) time binning. Despite the different methods adopted to estimate the integral flux, we observe a very good agreement between the two pipelines: the estimated flux values display differences between $20\%$ for the run-wise binning and less than $5\%$ for the weekly binning. Uncertainties on the estimated fluxes display deviations within $20\%$, with those computed by \gammapy about $10\%$ smaller than those estimated with \mars. The differences observed in the estimation of integral fluxes and their uncertainties are further investigated in \ref{sec:flux_detailed_comparison}, where we replicate the LC computation performed by \mars with \gammapy.

\subsubsection{Spectrum of the \crab observed under different moonlight conditions}
\label{sec:crab_moon_results}

\begin{figure}
    \centering
    \resizebox{\hsize}{!}{\includegraphics{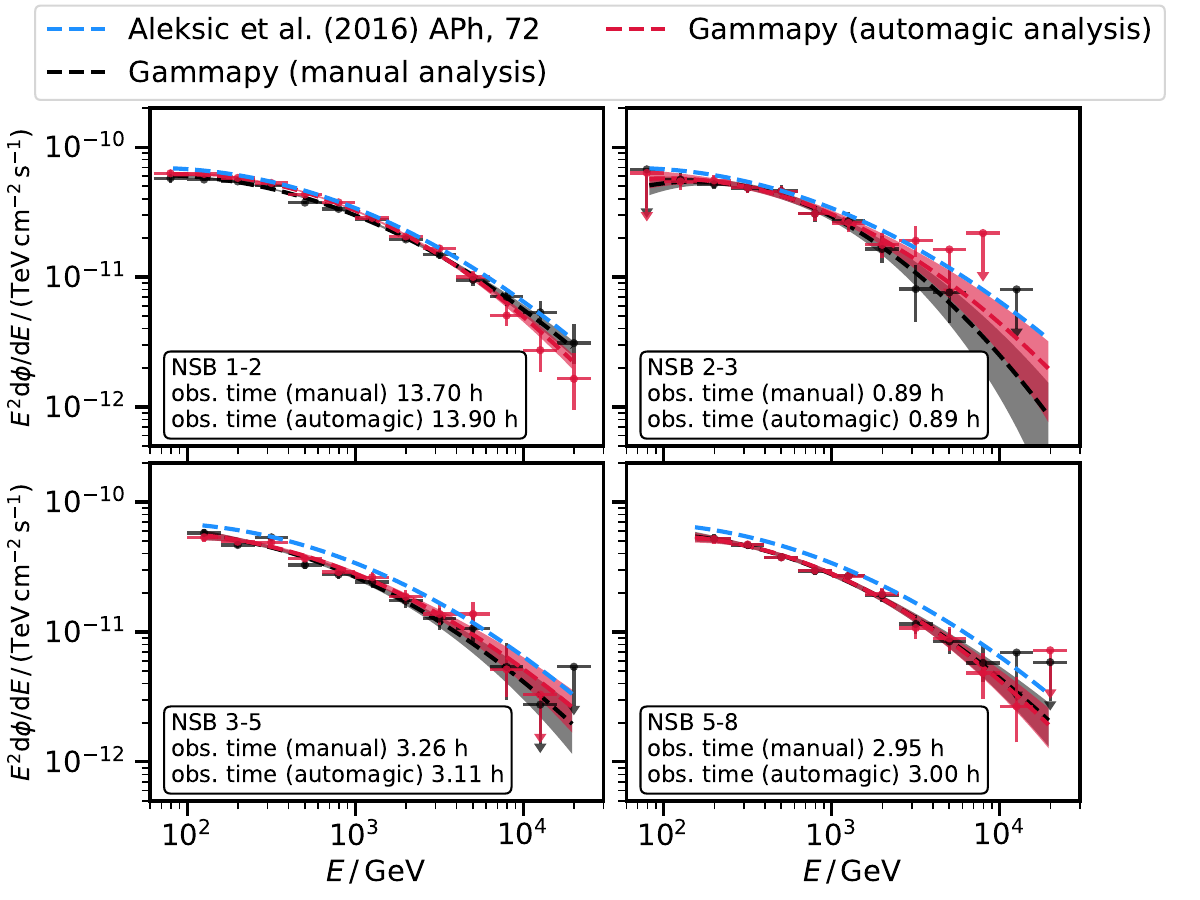}}
    \caption{SED of \crab obtained using a manual and an automated data selection and reduction. The quantity of data selected per each moon level by the two analyses is expressed through the data set effective time. Each NSB level indicates the intensity of the moon illumination.}
    \label{fig:crab_sed_moon}
\end{figure}

To validate the \automagic pipeline, we use the \crab observation taken under different moon conditions described in Sect.~\ref{sec:crab_moon_dataset} and compare the spectra estimated using the DL3 data produced both manually and automatically. We classify the data in NSB levels according to the direct current measured in the photomultipliers of the \magic~I telescope \citep[according to the prescriptions of Table~1 of][]{magic_performance_moonlight}. We consider NSB levels from 1 to 8 (in units of dark NSB level) and group them into four larger bins: NSB $1$-$2$, NSB $2$-$3$, NSB $3$-$5$, and NSB $5$-$8$. Fig.~\ref{fig:crab_sed_moon} shows the spectra estimated for the four different NSB bins, along with the quantity of data selected per each level, specified through the data set effective time. Both the manual and \automagic high-level analyses are performed with DL3 data and \gammapy. For what concerns the data selection, the manual and automatic procedures select and classify similar amounts of data in each NSB bin, with discrepancies of a few minutes due to different data selection procedures. We observe a very good agreement between the spectra obtained with the two approaches, for all the considered moon levels A detailed comparison of the estimated spectral parameters and their errors is provided in Fig.~\ref{fig:crab_parameters}. We notice that, despite the tuned image cleaning, in the bins with the highest NSB levels (3-5 and 5-8) the spectrum of the \crab remains slightly underestimated with respect to the reference. This behaviour is also observed in the moon performance study \citep[see Fig.~10 of][and the systematics evaluation therein]{magic_performance_moonlight}.

\subsubsection{Estimation of soft and hard gamma-ray spectra, \ebl absorption}
\label{sec:soft_hard_source}
\begin{figure}
    \centering
    \resizebox{\hsize}{!}{\includegraphics{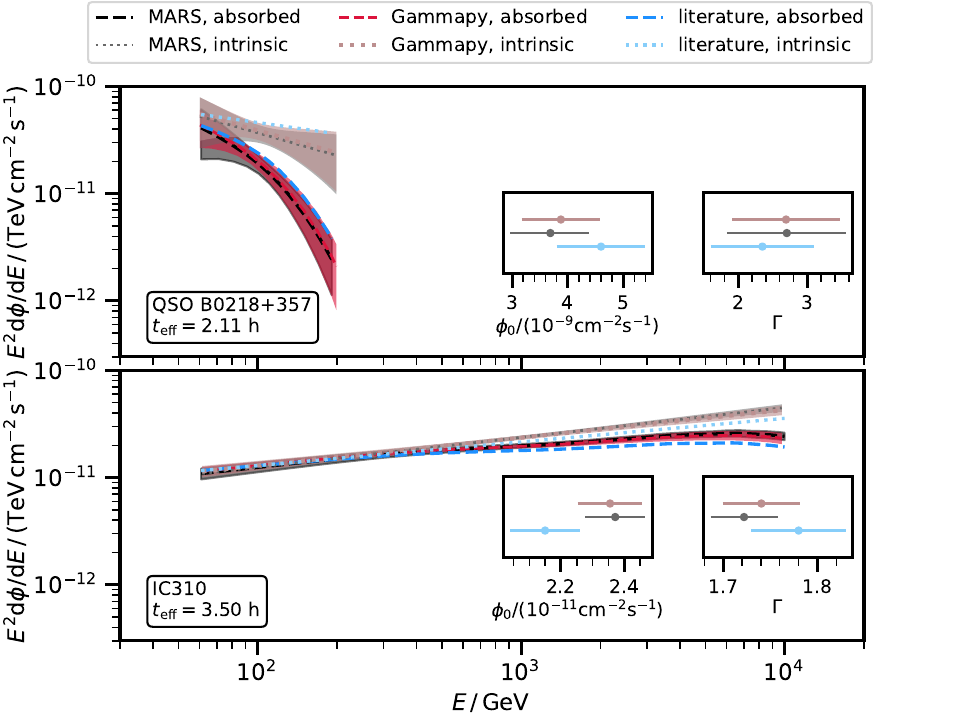}}
    \caption{SEDs of \qso (\textit{top}) and \ic (\textit{bottom}). Lighter colours represent the estimated spectra without the effect of the \ebl absorption. The inlets represent the estimated spectral parameters.}
    \label{fig:soft_hard_source_sed_comparison}
\end{figure}

We estimate the spectra of the jetted AGN \qso and \ic assuming a power law function
\begin{equation}
\frac{\diff \phi}{\diff E}(E; \phi_0, \Gamma, E_0) = \phi_0 \, \left( \frac{E}{E_0} \right)^{-\Gamma},
\label{eq:power_law}
\end{equation}
with the reference energy fixed to $100\,{\rm GeV}$ for \qso and to $1\,{\rm TeV}$ for \ic. We consider, in both software, the effect of the \ebl absorption according to the model of \cite{dominguez_2011}. Fig.~\ref{fig:soft_hard_source_sed_comparison} displays the results of the spectrum estimation for both sources. We observe a very good agreement between the two pipelines for both spectral types and for both distances (and absorption factors) considered.

\subsubsection{Light curve of a highly-variable source}
\label{sec:mrk_results}

\begin{figure*}
    \centering
    \includegraphics[width=17cm]{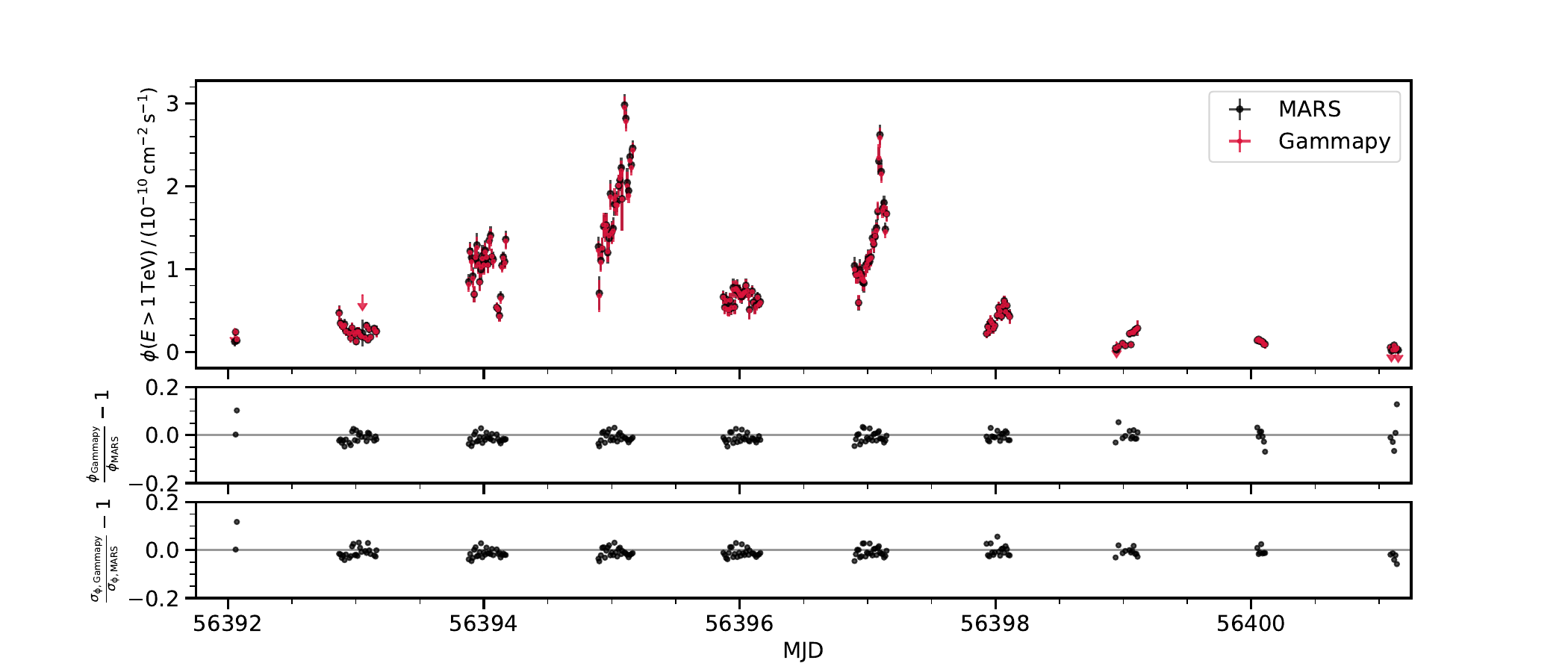}
    \caption{Run-wise LC of \mrk during April 2013 computed with \mars and \gammapy.}
    \label{fig:mrk421_lc}
\end{figure*}

We perform a light curve estimation using the \mrk data set described in Sect.~\ref{sec:mrk_dataset}. Remarking the different method to estimate the LC described in Sect.~\ref{sec:crab_multi_offset_results}, we observe a good agreement between the two software, as illustrated in Fig.~\ref{fig:mrk421_lc}, with the estimated flux values differing by $20\%$ and the estimated uncertainites showing a similar deviation. 

\subsubsection{Upper limits on the flux of a non-detected object}
\label{sec:m15_results}

\begin{figure}
    \resizebox{\hsize}{!}{\includegraphics{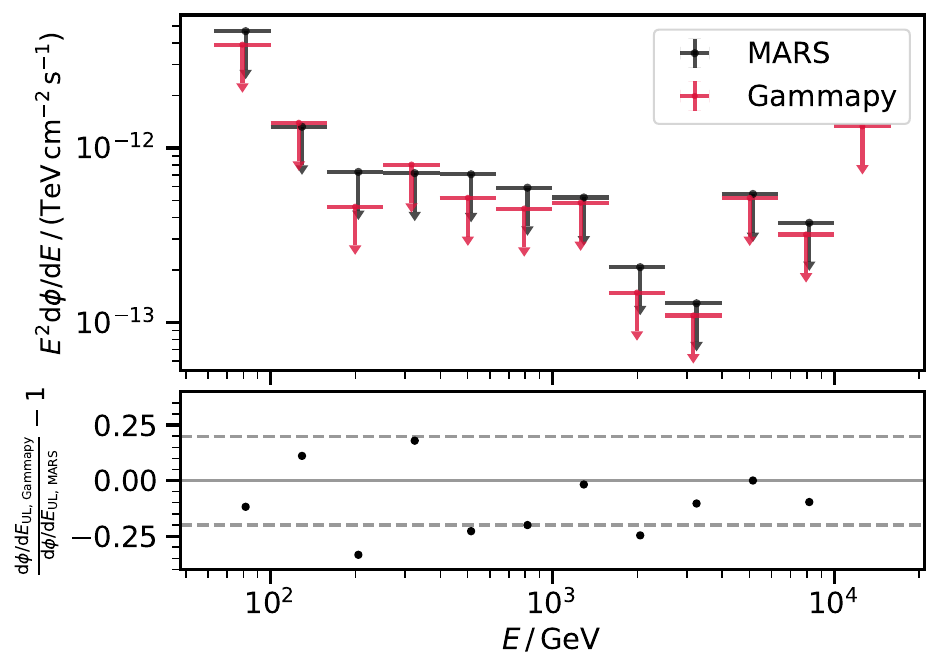}}
    \caption{UL on the gamma-ray flux of M15 computed with \mars and \gammapy.}
    \label{fig:m15_ul}
\end{figure}

As a last example of the estimations commonly performed in a one-dimensional analysis, we compute $95\%$ confidence level upper limits (ULs) on the gamma-ray flux of M15 using both \mars and \gammapy. By considering a likelihood ratio test based on Eq.~\ref{eq:likelihood}, \mars estimates a confidence level on the number of predicted counts $g$, that is then converted into a flux upper limit. For the likelihood profiling, the \cite{rolke_2001} method is applied. \gammapy computes instead the upper limit on the flux using the likelihood in Eq.~\ref{eq:likelihood} to estimate a one-sided confidence interval on the amplitude parameter, $\phi_0$. As illustrated in Fig.~\ref{fig:m15_ul} a good agreement between the two methods is found, with deviations below $30\%$. We remark that typically \mars considers a $30\%$ systematic uncertainty on the efficiency of the detected gamma rays (not applied in this particular comparison). The inclusion of this systematic uncertainty results in a $\sim20\%$ higher flux UL.

\section{Data availability}
\label{sec:data_availability}
As the purpose of this paper is not only to present the standardisation effort, but also to initiate the dissemination of the \magic standardised data, we make all the \crab DL3 data used in this article publicly available both via the \magic data centre at PIC\footnote{\url{https://opendata.magic.pic.es/}, under ``MAGIC DL3 Public Data Release 1''.} and via \texttt{zenodo} \citep{magic_dl3_pdr1}\footnote{\url{https://zenodo.org/records/11108474}}. Differently than previous releases of DL3 data \citep[e.g.][]{joint_crab}, we release the \crab data sets with a permissive license (Creative Commons Attribution, CC BY\footnote{\url{https://creativecommons.org/licenses/by/4.0/}}) that allows for their usage for scientific investigations.

\section{Conclusion}
\label{sec:conclusion}
In this paper, we presented the systematic effort to convert observations of the \magic telescopes to the standardised \gadf data format and to validate their analysis with the open-source software \gammapy. We considered six data sets representing different scientific and technical analysis cases. We focus our attention on the one-dimensional analysis and reproduce with the standardised data and \gammapy the computations commonly performed in such an analysis: estimation of a spectrum, of a light curve, and of upper limits on the flux in case of a non detection. In all the cases considered, even when different flux estimation methods are adopted, we observe a good agreement within $20\%$ between the results of the \magic proprietary analysis chain and the standardised data analysed with \gammapy (except upper limits where we observe deviations up to $35\%$). We demonstrate in \ref{sec:flux_detailed_comparison} that when replicating the same flux estimation algorithms of \mars in \gammapy, differences in the estimated values and uncertainties reduce to below $5\%$. As the realisation of any future \magic data legacy will require the systematic conversion of large amounts of data, we presented in this publication also the implementation of \automagic, a database-supported pipeline whose purpose is to produce standardised data automatising the process of data selection and reduction in case of different observing periods and conditions. We devoted part of the validation to cross-check the result of the automated pipeline against those of a manual analysis.
\par
Considering future prospects from the technical point of view, the next verification to be performed using the standardised data would be that of the spectro-morphological or three-dimensional analysis. A study of the gamma-ray background of the \magic telescopes, mandatory for such an analysis, is already underway \citep[see][]{bkg_simone} and we will present this validation in a follow-up publication. 
%In the meanwhile, having validated with this paper all the technical aspects of the point-like analysis, the \magic Collaboration can already provide standardised data that cover most of its observations and analysis cases.
%% BEFORE IN THE INTRODUCTION:
%The validation of the spectro-morphological, or three-dimensional, analysis is left to a follow-up publication that will also present a detailed study of the gamma-ray background of the MAGIC telescopes \citep[see the preliminary work in][]{bkg_simone}.
\par
From the point of view of the community, the validation conducted in this paper makes us confident that the standardised data correctly encapsulate the information in the proprietary one, and that the open-source analysis tools produce results consistent with those obtained with the proprietary software. This process already granted Collaboration members the possibility to adopt standardised data and open-source analysis tools for their analyses. In the hope to extend this possibility to the community in the near future, and to already encourage the exploration of \magic data, we make all the \crab standardised data used in this paper publicly available. This represents the first major release of \magic data to the public.
\par
This work represents the first milestone in the realisation of the \magic data legacy: future data releases will follow this publication, inaugurating the public scientific exploitation of two decades of observations. We remark that this scientific exploitation can be conducted in an accessible and fully-reproducible manner thanks to the efforts already devoted by the community to the data standardisation and to the development of open-source analysis tools. The production of \vhe gamma-ray data legacies, and the consequent possibility to combine decades of archival gamma-ray data, could deliver a closing statement from this current generation of instruments on several open questions in high-energy astrophysics. This would be of fundamental importance in deciding the most profitable scientific avenues to be pursued with the next-generation of observatories.

\section*{Author contributions}
\textbf{Cosimo Nigro}: conceptualisation, methodology, software, validation, formal analysis, resources, data curation, writing - original draft, writing - review \& editing, visualisation. \textbf{Lena Linhoff}: conceptualisation, software, validation, reosurces, data curation, writing - original draft. \textbf{Simone Mender}: software, validation, reosurces, data curation. \textbf{Jan Lukas Schubert}: software, validation, reosurces, data curation. \textbf{Maximilian Linhoff}: software, validation. \textbf{Julian Sitarek}, \textbf{Marcel Strzys}, \textbf{Ivana Batkovi{\'c}}, \textbf{Mireia Nievas Rosillo}: reosurces, data curation. The rest of the authors have contributed in one or several of the following ways: design, construction, maintenance and operation of the instrument(s) used to acquire the data; preparation and/or evaluation of the observation proposals; data acquisition, processing, calibration and/or reduction; production of analysis tools and/or related Monte Carlo simulations; overall discussions about the contents of the draft, as well as related refinements in the descriptions.

\section*{Acknowledgements}
We would like to acknowledge the early \magic data standardisation efforts by Tarek Hassan and Léa Jouvin. We would like to thank also Regis Terrier, Axel Donath, and all the \gammapy developers for the support they provided during the analysis development and validation.

We would like to thank the Instituto de Astrof\'{\i}sica de Canarias for the excellent working conditions at the Observatorio del Roque de los Muchachos in La Palma. The financial support of the German BMBF, MPG and HGF; the Italian INFN and INAF; the Swiss National Fund SNF; the grants PID2019-104114RB-C31, PID2019-104114RB-C32, PID2019-104114RB-C33, PID2019-105510GB-C31, PID2019-107847RB-C41, PID2019-107847RB-C42, PID2019-107847RB-C44, PID2019-107988GB-C22, PID2022-136828NB-C41, PID2022-137810NB-C22, PID2022-138172NB-C41, PID2022-138172NB-C42, PID2022-138172NB-C43, PID2022-139117NB-C41, PID2022-139117NB-C42, PID2022-139117NB-C43, PID2022-139117NB-C44 funded by the Spanish MCIN/AEI/ 10.13039/501100011033 and “ERDF A way of making Europe”; the Indian Department of Atomic Energy; the Japanese ICRR, the University of Tokyo, JSPS, and MEXT; the Bulgarian Ministry of Education and Science, National RI Roadmap Project DO1-400/18.12.2020 and the Academy of Finland grant nr. 320045 is gratefully acknowledged. This work was also been supported by Centros de Excelencia ``Severo Ochoa'' y Unidades ``Mar\'{\i}a de Maeztu'' program of the Spanish MCIN/AEI/ 10.13039/501100011033 (CEX2019-000920-S, CEX2019-000918-M, CEX2021-001131-S) and by the CERCA institution and grants 2021SGR00426 and 2021SGR00773 of the Generalitat de Catalunya; by the Croatian Science Foundation (HrZZ) Project IP-2022-10-4595 and the University of Rijeka Project uniri-prirod-18-48; by the Deutsche Forschungsgemeinschaft (SFB1491) and by the Lamarr-Institute for Machine Learning and Artificial Intelligence; by the Polish Ministry Of Education and Science grant No. 2021/WK/08; and by the Brazilian MCTIC, CNPq and FAPERJ.

%% bibliography
\bibliographystyle{elsarticle-harv}
\bibliography{biblio.bib} % your references Yourfile.bib

%% appendix starts here
\begin{appendix}

\section{Comparison of likelihood results for the \crab data}
\label{sec:likelihood_results_comparison}
\begin{figure}
    \centering
    \resizebox{\hsize}{!}{\includegraphics{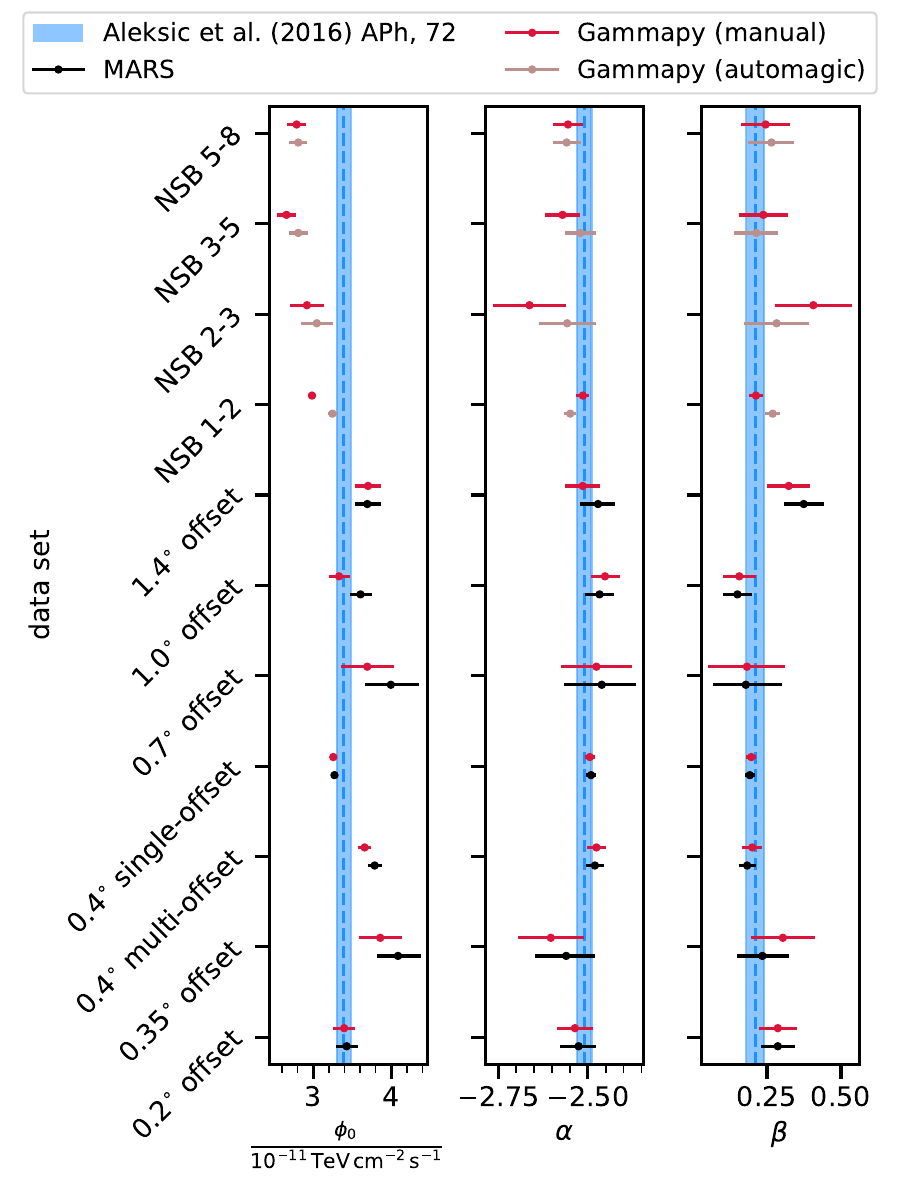}}
    \caption{Log-parabola (Eq.~\ref{eq:log_parabola}) parameters obtained from the likelihood fit performed with both software for all the \crab data sets in this paper. The values obtained in \cite{magic_performance} from the same data set are represented as a blue band for comparison.}
    \label{fig:crab_parameters}
\end{figure}

In Figure~\ref{fig:crab_parameters}, we illustrate the parameters obtained fitting the log-parabola spectrum to all \crab data in our validation sample, along with their errors. We also draw, for comparison, the parameters obtained in the performance paper \citep{magic_performance} using \mars and considering a sub sample of the single-offset data set.

\section{Detailed comparison of the integral flux computation with both software}
\label{sec:flux_detailed_comparison}

\begin{figure*}
    \centering
    \includegraphics[width=17cm]{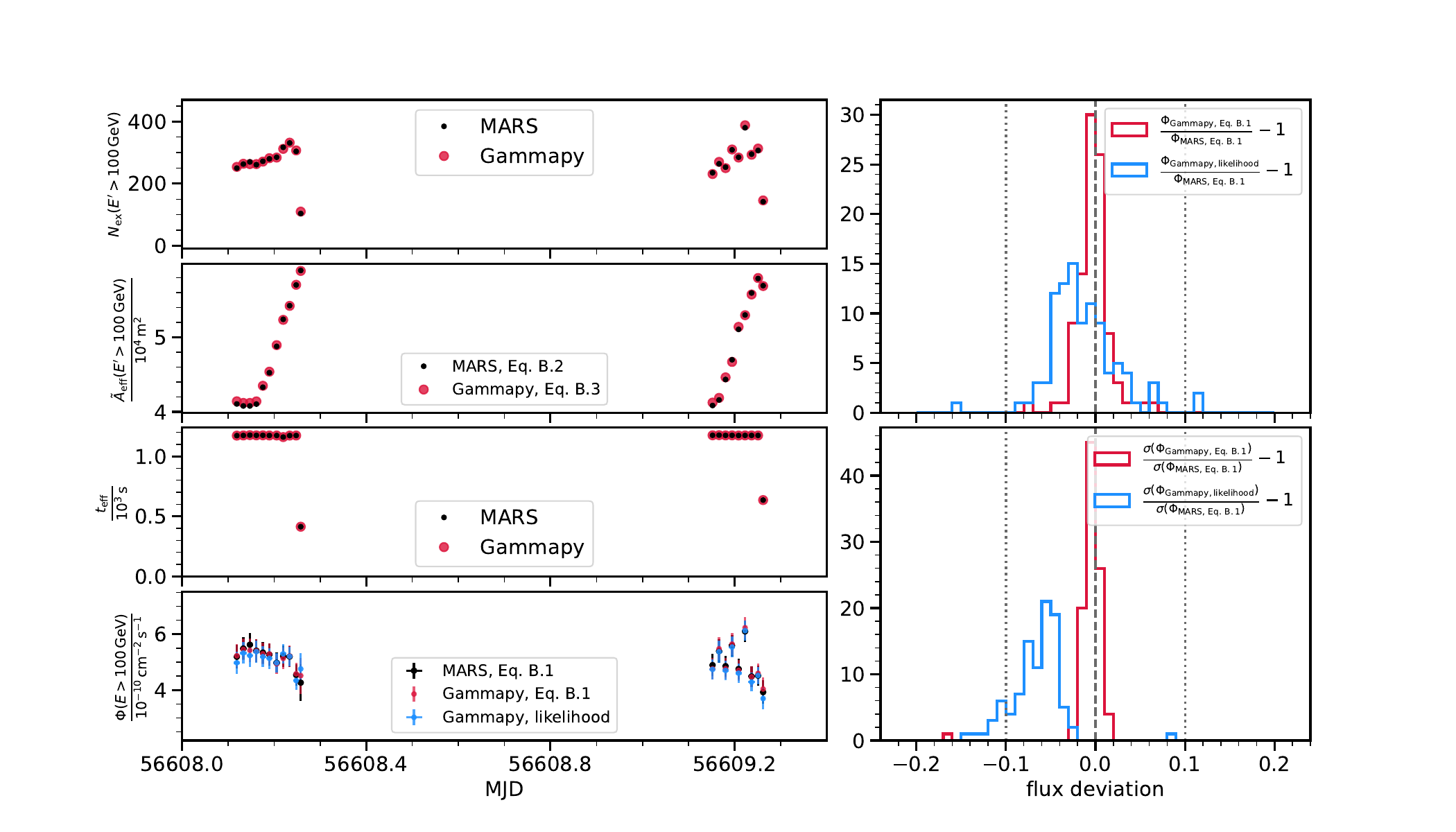}
    \caption{Replicating the \mars integral flux estimation with \gammapy. In the left panels, we compare the factors used for the integral flux estimation and the fluxes estimated with both software. For \gammapy, we consider both a re-implementation of \mars flux estimation (red) as well as its default likelihood-based estimation (blue). For simplicity, only two consecutive nights of the single-offset \crab data set are shown. In the right panels, we illustrate the deviations of the flux values and uncertainties estimated by \gammapy with both method (MARS-like and likelihood-based) from those estimated by \mars considering the whole \crab single-offset data set.}
    \label{fig:detailed_lc_comparison}
\end{figure*}

In order to investigate the systematic $10\%$ discrepancy observed between the uncertainties on the integral flux estimated by \gammapy and \mars (see Fig.~\ref{fig:crab_lc}), we replicate the integral flux computation performed by \mars using the \gammapy routines. Let $E_{\rm min}$ be the energy above which we would like to estimate the integral flux in a given time bin. The latter is computed by \mars as

\begin{equation}
\Phi(E > E_{\rm min}) = \frac{N_{\rm ex}(E' > E_{\rm min})}{\tilde{A}_{\rm eff}(E' > E_{\rm min})\,t_{\rm eff}},
\label{eq:mars_integral_flux}
\end{equation}

\noindent where $N_{\rm ex}(E' > E'_{\rm min})$ is the total number of excess events ($N_{\rm on} - \alpha N_{\rm off}$, see Sec.~\ref{sec:spectral_analysis}) with estimated energy above $E_{\rm min}$, $t_{\rm eff}$ is the effective time, and $\tilde{A}_{\rm eff}(E' > E_{\rm min})$ is an effective area modified to take into account the energy dispersion. It is computed as 
\begin{equation}
\tilde{A}_{\rm eff}(E' > E_{\rm min}) = \frac{N_{\gamma, {\rm after\,cuts}}(E' > E_{\rm min})}{N_{\gamma, {\rm simulated}}(E > E_{\rm min})} A_{\rm simulated},
\label{eq:aeff_poor_man_unfold_mars}
\end{equation}
where in the numerator we have the total number of simulated events with estimated energy above $E_{\rm min}$ that survive the analysis cuts and in the denominator we have the total number of simulated events with true energy above $E_{\rm min}$. $A_{\rm simulated}$ represents the total simulated area. Obtaining the quantities in Eq.~\ref{eq:aeff_poor_man_unfold_mars} would require having access to the information relative to the individual simulated events, which is no longer available at the reduction level of the DL3 data. We therefore make use of the \irf components stored in the DL3 files to obtain such an effective area. Given an assumed spectral model, $\frac{\diff \phi}{\diff E}$ that we take to be the best-fit model obtained from the likelihood maximisation (Eq.~\ref{eq:likelihood}), we compute the effective area above a certain estimated energy as
\begin{equation}
\tilde{A}_{\rm eff}(E' > E_{\rm min}) = 
\frac{\displaystyle\frac{\diff N}{\diff t}(E' > E_{\rm min})}{\displaystyle\int_{E > E_{\rm min}} \frac{\diff \phi}{\diff E} \diff E},
\label{eq:aeff_poor_man_unfold_gammapy}
\end{equation}
where $\frac{\diff N}{\diff t}(E' > E_{\rm min})$ is the rate of gamma rays with estimated energies $E' > E_{\rm min}$. The rate in a given true energy bin can be obtained as
\begin{equation}
\frac{\diff N}{\diff t}(E_{i}) = \int_{\Delta E_{i}} A_{\rm eff}(E) \frac{\diff \phi}{\diff E}\diff E
\label{eq:rate_true_energy}
\end{equation}
where $A_{\rm eff}(E)$ is the effective area as a function of true energy provided in the \gadf-compliant \irf components and $i$ is an index running over the true energy bins, while $\Delta E_{i}$ the true energy bin width. To obtain the rate above an estimated energy, we multiply Eq.~\ref{eq:rate_true_energy} with the migration matrix $R(E'|E)$
\begin{equation}
\frac{\diff N}{\diff t}(E' > E_{\rm min}) = \sum_{j\,:\,E' > E_{\rm min}} \frac{\diff N}{\diff t}(E_{i}) R_{ij},
\label{eq:rate_estimated_energy}
\end{equation}
where now $j$ is an index that runs over the estimated energy bins and we have represented the energy migration as an $i \times j$ matrix. Using the \gammapy routines, we can compute the effective area in Eq.~\ref{eq:aeff_poor_man_unfold_gammapy} from the \irf components in the DL3 files ($A_{\rm eff}(E)$ and $R(E'|E)$). The integral flux is then obtained, according to Eq.~\ref{eq:mars_integral_flux}, by considering the total number of excess events counted in an observation and its effective time. For the uncertainty on the flux we neglect any uncertainty on the effective time and effective area and consider only the uncertainty on the number of excess events, obtained as $\sigma(N_{\rm ex}) = \sqrt{N_{\rm on} + \alpha^2 N_{\rm off}}$. 

In the left panels of Fig.~\ref{fig:detailed_lc_comparison} we illustrate the factors in Eq.~\ref{eq:mars_integral_flux} obtained with both software for two consecutive nights in the single-offset \crab data set. In the bottom panel, we illustrate the fluxes obtained by MARS and by \gammapy, with the latter using both Eq.~\ref{eq:mars_integral_flux} and the result of the likelihood maximisation. In the right panels of the figure, we illustrate the deviations of the flux values and uncertainties estimated by \gammapy from those estimated by \mars, considering the whole \crab single-offset data set. As one can see, when adopting the same method as \mars, flux values and uncertainties estimated with \gammapy differ by less than $5\%$ from those estimated by \mars. We are thus confident that the $10\%$ underestimation of the integral flux uncertainty observed when using \texttt{gammapy}'s default likelihood routine is due to the different approaches adopted for the estimation and not to any issue with the standardised DL3 data.

\end{appendix}

\end{document}